\documentclass{article}%
\usepackage{amsmath}%
\usepackage{amsfonts}%
\usepackage{amssymb}%
\begin{document}

\author{Vladimir K. Petrov\thanks{ E-mail address: vkpetrov@yandex.ru}}
\title{{\LARGE Some remarks on cyclization and periodicity}}
\date{\textit{N. N. Bogolyubov Institute for Theoretical Physics}\\
\textit{\ National Academy of Sciences of Ukraine}\\
\textit{\ 252143 Kiev, Ukraine. 10.11.2002}}
\maketitle

\begin{abstract}
Conditions when application of \textit{cyclization} procedure does not lead to
the periodic function are specified. Sufficient conditions under which
obtained function is periodic are stated.

\end{abstract}

\section{Introduction}

Let us put in correspondence to any function $F\left(  \varphi\right)  $
function
\begin{equation}
\widetilde{F}\left(  \varphi\right)  =\sum_{n=-\infty}^{\infty}F\left(
\varphi+2\pi n\right) \label{per}%
\end{equation}

This procedure $\left(  \ref{per}\right)  $, which we call
\textit{cyclization}, is used not infrequently (see e.g. \cite{Schwartz}), but
as a rule is applied to such functions $F\left(  \varphi\right)  $ that
decrease quite rapidly with $\varphi\rightarrow\pm\infty$ and $\widetilde
{F}\left(  \varphi\right)  $ turns out to be periodic. In this paper we
consider cases when the above condition is not true and $\widetilde{F}\left(
\varphi\right)  \neq\widetilde{F}\left(  \varphi+2\pi\right)  $.

It is worth noting from the outset that since shift $\widetilde{F}\left(
\varphi\right)  \rightarrow$ $\widetilde{F}\left(  \varphi+2\pi\right)  $
leads to some sort of interchange of summations, function $\widetilde
{F}\left(  \varphi\right)  $ may appear to be aperiodic, if series $\left(
\ref{per}\right)  $ is not absolutely convergent.

Indeed, if we take as an example
\begin{equation}
F\left(  \varphi\right)  =\ln\left(  1+\frac{\alpha}{\varphi}\right)
-\frac{\alpha}{\varphi},\label{ex}%
\end{equation}
then after cyclization we get
\begin{equation}
\widetilde{F}\left(  \varphi\right)  =\sum_{n=-\infty}^{\infty}\ln\left(
\left(  1+\frac{\alpha}{\varphi+2\pi n}\right)  \exp\left\{  -\frac{\alpha
}{\varphi+2\pi n}\right\}  \right)
\end{equation}
which, taking into account (\cite{pbm} 6.2.3.$\left(  8\right)  $) may be
written as
\begin{equation}
\widetilde{F}\left(  \varphi\right)  =\frac{\alpha}{2\pi}\allowbreak
\operatorname{Psi}\left(  \frac{\varphi}{2\pi}\right)  +\ln\frac{\Gamma\left(
\frac{\varphi}{2\pi}\right)  }{\Gamma\left(  \frac{\varphi+\alpha}{2\pi
}\right)  }%
\end{equation}

It is easy to check that
\begin{equation}
\widetilde{F}\left(  \varphi+2\pi\right)  -\widetilde{F}\left(  \varphi
\right)  =\frac{\alpha}{\varphi}-\ln\left(  1+\frac{\alpha}{\varphi}\right)
\neq0
\end{equation}

In many interesting cases series in $\left(  \ref{per}\right)  $ is divergent
in ordinary sense, then to compute $\widetilde{F}\left(  \varphi\right)  $ it
is rational to apply distribution theory where rigorous methods is developed
for summing extensive class of such series
\cite{Schwartz,gel-shil,vladimirov,bremermann}. In particular, such methods
allow to compute integrals of tempered distributions and series which terms
coincide with some tempered function at integer values of the argument
\cite{Schwartz}.

Recall that $F\left(  \varphi\right)  $ is tempered distribution, if and only
if it is finite order derivative of some tempered function. Continuous
$f\left(  \varphi\right)  $ function is called a tempered one
\cite{bremermann}, if exists some $\sigma$ that
\begin{equation}
\left|  f\left(  \varphi\right)  \right|  <\left|  \varphi\right|  ^{\sigma
};\qquad\left|  \varphi\right|  \rightarrow\infty.\label{cond_t}%
\end{equation}

It is important that every periodic distribution $\widetilde{F}_{\left(
p\right)  }\left(  \varphi\right)  $ is tempered \cite{donoghue} and Fourier
transform of any tempered distribution is tempered distribution as well
\cite{bremermann}.

It is known that any tempered distribution $F\left(  \varphi\right)  $ or
$\widetilde{F}_{\left(  p\right)  }\left(  \varphi\right)  $ may be presented
as (see e.g. \cite{vladimirov}).
\begin{equation}
F\left(  \varphi\right)  =F_{+}\left(  \varphi+i\varepsilon\right)
-F_{-}\left(  \varphi-i\varepsilon\right) \label{vla-i}%
\end{equation}
and
\begin{equation}
\widetilde{F}_{\left(  p\right)  }\left(  \varphi\right)  =\widetilde{F}%
_{+}\left(  \varphi+i\varepsilon\right)  -\widetilde{F}_{-}\left(
\varphi-i\varepsilon\right)  ;\label{vla}%
\end{equation}
where $\widetilde{F}_{\pm}\left(  \varphi\right)  $ and $F_{\pm}\left(
\varphi\right)  $ are analytical functions in upper/lower complex half-plane
$\varphi$ and $\varepsilon$ is routine positive infinitesimal parameter.

It may be easily checked, if to choose
\begin{equation}
F_{+}\left(  \varphi\right)  \equiv\int_{0}^{\infty}F_{t}e^{i\varphi
t}dt;\qquad F_{-}\left(  \varphi\right)  \equiv-\int_{-\infty}^{0}%
F_{t}e^{i\varphi t}dt
\end{equation}
and
\begin{equation}
\widetilde{F}_{+}\left(  \varphi\right)  \equiv\sum_{n=0}^{\infty}%
F_{n}e^{i\varphi n};\qquad\widetilde{F}_{-}\left(  \varphi\right)  \equiv
-\sum_{n=-\infty}^{-1}F_{n}e^{i\varphi n}%
\end{equation}
then expression $\left(  \ref{vla-i}\right)  $ will coincide with an
Abel-Poisson regularization of Fourier integral
\begin{subequations}
\begin{equation}
F\left(  \varphi\right)  =\int_{-\infty}^{\infty}F_{t}e^{it\varphi
}dt\rightarrow\int_{-\infty}^{\infty}F_{t}e^{it\varphi-\varepsilon\left|
t\right|  }dt\label{f-i}%
\end{equation}
and $\left(  \ref{vla}\right)  $ with an Abel-Poisson regularization of
Fourier series \
\end{subequations}
\begin{equation}
\widetilde{F}_{\left(  p\right)  }\left(  \varphi\right)  \equiv
\sum_{n=-\infty}^{\infty}F_{n}e^{i\varphi n}\rightarrow\sum_{n=-\infty
}^{\infty}F_{n}e^{i\varphi n-\varepsilon\left|  n\right|  }\label{A-P}%
\end{equation}

Abel-Poisson regularization provide continuous convergence Fourier integrals
$\left(  \ref{f-i}\right)  $ and series $\left(  \ref{A-P}\right)  $ in a case
when $F_{t}$ is tempered distribution and Fourier coefficients $F_{n}$
coincide with values of some tempered distribution at integer $t=n$.

In particular, due to continuous convergence of regularized integrals, the
order of integration in double integrals may be changed and it may be shown,
that the transform inverse to $\left(  \ref{f-i}\right)  $ is
\begin{subequations}
\begin{equation}
F_{t}=\frac{1}{2\pi}\int_{-\infty}^{\infty}F\left(  \varphi\right)
\exp\left\{  -i\varphi t-\varepsilon\left\vert \varphi\right\vert \right\}
d\varphi,\label{f-t}%
\end{equation}
Furthermore, since regularized\footnote{Here $\varepsilon^{\prime}$ is
positive infinitesimal parameter independent of $\varepsilon$.} sums and
integrals in
\end{subequations}
\begin{align}
& \sum_{n=-\infty}^{\infty}F\left(  \varphi-2\pi n\right) \nonumber\\
& =\sum_{n=-\infty}^{-1}\int_{-\infty}^{0}e^{i\left(  \varphi-2\pi
n-i\varepsilon\right)  \left(  t+i\varepsilon^{\prime}\right)  }F_{t}%
dt+\sum_{n=-\infty}^{-1}\int_{0}^{\infty}e^{i\left(  \varphi-2\pi
n+i\varepsilon\right)  \left(  t+i\varepsilon^{\prime}\right)  }%
F_{t}dt+\nonumber\\
& \sum_{n=0}^{\infty}\int_{-\infty}^{0}e^{i\left(  \varphi-2\pi n-i\varepsilon
\right)  \left(  t-i\varepsilon^{\prime}\right)  }F_{t}dt+\sum_{n=0}^{\infty
}\int_{0}^{\infty}e^{i\left(  \varphi-2\pi n+i\varepsilon\right)  \left(
t-i\varepsilon^{\prime}\right)  }F_{t}dt
\end{align}
are continuously convergent it allows to change the order of integration and
summation, thus taking into account the Poisson relation
\begin{equation}
\sum_{n=-\infty}^{\infty}e^{i2\pi nm}=\sum_{n=-\infty}^{\infty}\delta\left(
n-m\right) \label{del-P}%
\end{equation}
one can easily find that application of the procedure $\left(  \ref{per}%
\right)  $\ to regularized integral $\left(  \ref{f-i}\right)  $\ gives
\begin{equation}
\widetilde{F}\left(  \varphi\right)  =\int_{-\infty}^{\infty}\sum_{n=-\infty
}^{\infty}e^{i\left(  \varphi-2\pi n\right)  t}F_{t}dt=\sum_{m=-\infty
}^{\infty}\int_{-\infty}^{\infty}e^{i\varphi t}\delta\left(  t-m\right)
F_{t}dt\label{se}%
\end{equation}

In a case when for all integer $n$ function $F_{t}$ obeys the condition \
\begin{equation}
F_{n}=\lim_{t\rightarrow n}F_{t}<\infty\label{fin}%
\end{equation}
series $\left(  \ref{se}\right)  $ coincides with $\left(  \ref{A-P}\right)  $
and $F_{n}$\ can be considered Fourier coefficients which may be computed, as
it follows from $\left(  \ref{per}\right)  $ and $\left(  \ref{f-t}\right)  $,
from the ordinary relation
\begin{equation}
F_{n}=\frac{1}{2\pi}\int_{-\pi}^{\pi}\widetilde{F}\left(  \varphi\right)
\exp\left\{  -i\varphi n\right\}  d\varphi\label{Fn}%
\end{equation}

For $F_{t}$ to exist function $F\left(  \varphi\right)  $ should be only
locally summable rather than summable. However, if $F\left(  \varphi\right)  $
is summable, its Fourier transform $F_{t}$ is bounded for any real $t$
\cite{Schwartz}.

It is worth to note, that too rigid constraint on class function to which
$F\left(  \varphi\right)  $ belongs, may lead to quite trivial class
function\ $F_{t}$\ and, consequently, trivial $\widetilde{F}\left(
\varphi\right)  $. Indeed, let $F\left(  \varphi\right)  $\ be infinitely
differentiable with compact support (i.e. belongs to Schwartz space $\left(
D\right)  $), then, according to \cite{gel-shil}, its Fourier transform
$F_{t}$\ belongs to space of entire functions $\left(  Z\right)  $. If in
addition
\begin{equation}
\lim_{\left\vert t\right\vert \rightarrow\infty}\frac{\ln\left\vert
F_{t}\right\vert }{\left\vert t\right\vert }=0
\end{equation}
for any direction in the complex plane $t$, and $F_{t}$\ is bounded for all
integer $t=n$, then, according Polya theorem (see e.g. \cite{pal-wein}),
$F_{n}=const$.

It is evident that, if condition $\left(  \ref{fin}\right)  $ is fulfilled,
function $\widetilde{F}\left(  \varphi\right)  $ given by $\left(
\ref{per}\right)  $ may be presented as regularized Fourier series $\left(
\ref{A-P}\right)  $ and consequently is periodic.

\section{Singular $F_{t}$}

In a case when function $F\left(  \varphi\right)  $ is not summable, Fourier
transform $F_{t}$ may appear to be singular at integer $t=n$, as in already
considered example $\left(  \ref{ex}\right)  $ where $F_{t}$, computed with (
\cite{brych-prud} 402), may be written as
\begin{equation}
F_{t}=\left(  e^{-i\alpha t}-1\right)  \left(  \pi\left(  2\allowbreak
\operatorname{Psi}\left(  1\right)  +1\right)  \delta\left(  t\right)
-\frac{i}{t}-\frac{\pi}{\left\vert t\right\vert }\right)  -i\pi\alpha
\operatorname*{signum}\left(  t\right)
\end{equation}

The case of singular $F_{t}$ needs special consideration. As it is seen from
$\left(  \ref{se}\right)  $, to obtain $\widetilde{F}\left(  \varphi\right)  $
we have to compute the product of two distributions: $\delta\left(
t-n\right)  $ and $F_{t}$. Unfortunately, the product of arbitrary
distributions is not defined (see e.g. \cite{Schwartz}), so we shall confine
ourself to the consideration of well-defined products, namely the case when
singularity of $F_{t}$ at some integer $t=k$ is a pole of order $m$. More
general case will be considered elsewhere. To be specific we consider $k=0$
i.e. $F_{t}\sim t^{-m}$ for $t\rightarrow0$.

Let us take as an example $F\left(  \varphi\right)  =\operatorname*{signum}%
\left(  \varphi\right)  $. Making allowance for (\cite{brych-prud} 13), one
may write
\begin{equation}
F\left(  \varphi\right)  =\operatorname*{signum}\left(  \varphi\right)
=\frac{1}{\pi i}\int_{-\infty}^{\infty}e^{it\varphi}t^{-1}dt
\end{equation}
and in this case
\begin{equation}
F_{t}=\frac{1}{\pi i}t^{-1}%
\end{equation}
is infinite at $t=0$.

It is easy to check that
\begin{equation}
\widetilde{F}\left(  \varphi\right)  =\sum_{n=-\infty}^{\infty}%
\operatorname*{signum}\left(  \varphi+2\pi n\right)  =\frac{1}{\pi i}%
\sum_{m=-\infty}^{\infty}\int_{-\infty}^{\infty}\delta\left(  m-t\right)
e^{it\varphi}t^{-1}dt
\end{equation}
which leads to
\begin{equation}
\widetilde{F}\left(  \varphi\right)  =\widetilde{F}_{\left(  p\right)
}\left(  \varphi\right)  +\widetilde{F}_{\left(  a\right)  }\left(
\varphi\right) \label{log}%
\end{equation}
where the first term is evidently periodic due to explicit invariance under
$\varphi\rightarrow\varphi+2\pi$%
\begin{equation}
\widetilde{F}_{\left(  p\right)  }\left(  \varphi\right)  =\sum_{m\neq0}%
\frac{e^{im\varphi}}{m}=-\frac{1}{\pi i}\ln\frac{1-e^{i\varphi}}%
{1-e^{-i\varphi}}%
\end{equation}
Since a logarithm is a many-valued function, we get the ambiguous result
\begin{equation}
-\frac{1}{\pi i}\ln\frac{1-e^{i\varphi}}{1-e^{-i\varphi}}=-\frac{1}{\pi i}%
\ln\left(  -e^{i\varphi}\right)  =-\frac{1}{\pi}\left(  \left(  \varphi\pm
\pi\right)  +2\pi n_{\varphi}\right)
\end{equation}
where $n_{\varphi}$ is an integer number, which numbers the branches of the logarithm.

To reduce ambiguity, we choose, as it is generally done, among the infinite
set of logarithm branches, the principal one
\begin{equation}
\frac{1}{\pi i}\sum_{m\neq0}\frac{e^{im\varphi}}{m}=-\frac{\left(  \varphi
\pm\pi\right)  _{\operatorname{mod}2\pi}}{\pi};\label{lo}%
\end{equation}
where the periodic function $\left(  x+2\pi\right)  _{\operatorname{mod}2\pi
}=x_{\operatorname{mod}2\pi}$ is defined in the interval $-\pi<x<\pi$ as
$x_{\operatorname{mod}2\pi}=x$ .

The second term in $\left(  \ref{log}\right)  $%
\begin{equation}
\widetilde{F}_{\left(  a\right)  }\left(  \varphi\right)  =\frac{1}{\pi i}%
\int_{-\infty}^{\infty}e^{it\varphi}\delta\left(  t\right)  t^{-1}dt
\end{equation}
has the singular integrand that with the help of a simple equation
\begin{equation}
\delta\left(  t\right)  t^{-n}n!=\left(  -1\right)  ^{n}\delta^{\left(
n\right)  }\left(  t\right)
\end{equation}
may be rewritten as
\begin{equation}
\widetilde{F}_{\left(  a\right)  }\left(  \varphi\right)  =-\frac{1}{\pi
i}\int_{-\infty}^{\infty}e^{it\varphi}\delta^{\left(  1\right)  }\left(
t\right)  dt=\frac{\varphi}{\pi}%
\end{equation}
This term is evidently aperiodic.

Taking into account%

\begin{equation}
\frac{\varphi-\varphi_{\operatorname{mod}2\pi}}{2\pi}=\left\lfloor
\frac{\varphi-\pi}{2\pi}\right\rfloor +1=\left\lfloor \frac{\varphi+\pi}{2\pi
}\right\rfloor
\end{equation}
where $\left\lfloor x\right\rfloor $ stands for integer part of $x$ (called
also $Entier\left(  x\right)  $), one may finally write%

\begin{equation}
\sum_{n=-\infty}^{\infty}\operatorname*{signum}\left(  \varphi+2\pi n\right)
=2\left\lfloor \frac{\varphi}{2\pi}\right\rfloor +1
\end{equation}

A more general case may be considered, making allowance (\cite{brych-prud}
22)
\begin{equation}
\left(  -1\right)  ^{m}\frac{\left(  2m\right)  !}{\pi i}\int_{-\infty
}^{\infty}e^{it\varphi}t^{-2m-1}dt=\left\vert \varphi\right\vert
^{2m}\operatorname*{signum}\left(  \varphi\right)  \equiv F\left(
\varphi\right)
\end{equation}
that leads to
\begin{align}
\widetilde{F}\left(  \varphi\right)   & =\sum_{n=-\infty}^{\infty}\left\vert
\varphi+2\pi n\right\vert ^{2m}\operatorname*{signum}\left(  \varphi+2\pi
n\right) \\
& =\left(  -1\right)  ^{m}\frac{\left(  2m\right)  !}{\pi i}\sum_{n=-\infty
}^{\infty}\int_{-\infty}^{\infty}e^{it\left(  \varphi+2\pi n\right)
}t^{-2m-1}dt\nonumber
\end{align}
so one may write
\begin{equation}
\widetilde{F}\left(  \varphi\right)  =\frac{\varphi^{2m+1}}{\left(
2m+1\right)  \pi}+\frac{\left(  2\pi\right)  ^{2m+1}}{\pi}\left(  2m\right)
!\sum_{k\neq0}\frac{e^{ik\varphi}}{\left(  2\pi ik\right)  ^{2m+1}}%
\end{equation}

Recall that (see e.g. \cite{bateman} 1.13$\left(  11\right)  $)
\begin{equation}
-n!\sum_{k\neq0}\frac{\exp\left\{  ik\varphi\right\}  }{\left(  2\pi
ik\right)  ^{n}}=B_{n}\left(  \frac{\varphi}{2\pi}\right)  ;\label{F-ser}%
\end{equation}
where $B_{n}\left(  x\right)  $ is Bernoulli polynomial of order $n$.
Unfortunately, expression $\left(  \ref{F-ser}\right)  $ is true only for
$0<\varphi<2\pi$ \cite{bateman}. One may see, however, that an extension of
$\left(  \ref{F-ser}\right)  $ on negative $\varphi$ may be easily done by the
substitution $\varphi\rightarrow\varphi+2\pi$, which doesn't effect explicitly
periodic right side of $\left(  \ref{F-ser}\right)  $, but transforms
condition $0<\varphi<2\pi$ into $-2\pi<\varphi<0$, and we may finally write
\begin{align}
& \sum_{n=-\infty}^{\infty}\left\vert \varphi+2\pi n\right\vert ^{2m}%
\operatorname*{signum}\left(  \varphi+2\pi n\right) \\
& =\frac{\varphi^{2m+1}}{\left(  2m+1\right)  \pi}-\frac{\left(  2\pi\right)
^{2m+1}}{\left(  2m+1\right)  \pi}B_{2m+1}\left(  \frac{\varphi
_{\operatorname{mod}2\pi}+2\pi\theta\left(  -\varphi_{\operatorname{mod}2\pi
}\right)  }{2\pi}\right)  .\nonumber
\end{align}

Proceeding as in previous case and taking into account (\cite{brych-prud} 21)
\begin{equation}
\left\vert \varphi\right\vert ^{2m-1}=\left(  -1\right)  ^{m}\frac{\left(
2m-1\right)  !}{\pi}\int_{-\infty}^{\infty}e^{it\varphi}\frac{1}{t^{2m}}dt
\end{equation}
we may compute
\begin{align}
\sum_{n=-\infty}^{\infty}\left\vert \varphi+2\pi n\right\vert ^{2m-1}  &
=\left(  -1\right)  ^{m}\frac{\left(  2m-1\right)  !}{\pi}\sum_{k\neq
0}e^{ik\varphi}k^{-2m}\\
& +\left(  -1\right)  ^{m}\frac{\left(  2m-1\right)  !}{\pi}\int_{-\infty
}^{\infty}e^{it\varphi}\delta\left(  t\right)  t^{-2m}dt\nonumber
\end{align}
that leads to%

\[
\sum_{n=-\infty}^{\infty}\left\vert \varphi+2\pi n\right\vert ^{2m-1}%
=-\frac{\left(  2\pi\right)  ^{2m-1}}{m}B_{2m}\left(  \frac{\varphi
_{\operatorname{mod}2\pi}}{2\pi}+\theta\left(  -\varphi\right)  \right)
+\frac{\varphi^{2m}}{2\pi m}%
\]

In particular for $m=1$ we get
\[
\widetilde{F}\left(  \varphi\right)  =\sum_{n=-\infty}^{\infty}\left|
\varphi+2\pi n\right|  =\frac{\varphi^{2}}{2\pi}-\frac{\varphi
_{\operatorname{mod}2\pi}^{2}}{2\pi}+\left|  \varphi_{\operatorname{mod}2\pi
}\right|  -\frac{\pi}{3}\allowbreak.
\]

It should be noted that for some functions $F\left(  \varphi\right)  $ after
lifting Abel-Poisson regularization $\widetilde{F}\left(  \varphi\right)  $
'converges' to infinity. In particular, for $F\left(  \varphi\right)
=\varphi^{m}e^{i\alpha\varphi}$%
\begin{equation}
\widetilde{F}\left(  \varphi\right)  =\sum_{n=-\infty}^{\infty}\left(
\varphi+2\pi n\right)  ^{m}e^{i\alpha\left(  \varphi+2\pi n\right)  }%
=i^{-m}\sum_{n=-\infty}^{\infty}e^{in\varphi}\delta^{\left(  m\right)
}\left(  \alpha-n\right)
\end{equation}
We see that for noninteger $\alpha$ function $\widetilde{F}\left(
\varphi\right)  =0$ whereas for integer $\alpha=k$ one gets $\widetilde
{F}\left(  \varphi\right)  =\left(  -\varphi\right)  ^{m}e^{ik\varphi}%
\delta\left(  0\right)  $

\section{Conclusions}

This paper is an attempt to reveal a mechanism that actualizes a known
connection between discretization of Fourier variables and periodicity in the
associated space.

It is shown that the application of procedure $\left(  \ref{per}\right)  $ to
function $F\left(  \varphi\right)  $ doesn't lead to the periodicity of
function $\widetilde{F}\left(  \varphi\right)  $ if Fourier transform of
$F\left(  \varphi\right)  $ has a pole at some integer value of its argument.
If, on the contrary, $F_{t}$ is finite for all integer $t=n$, function
$\widetilde{F}\left(  \varphi\right)  $ is periodic.

\end{document}